\documentclass[aps,prl,twocolumn]{revtex4}
\usepackage{amsmath,bm}
\usepackage{graphicx}

\newcommand{\avg}[1]{\langle #1 \rangle}
\newcommand{\var}[1]{\mathrm{Var}\left( #1 \right)}

\newcommand{\rtHz}{\sqrt{\text{Hz}}}

\begin{document}

\title{Quantum noise limited and entanglement-assisted magnetometry}

\author{W. Wasilewski}
\affiliation{Niels Bohr Institute, Danish Quantum Optics Center  QUANTOP, Copenhagen University, Blegdamsvej 17, 2100 Copenhagen, Denmark}
\author{K. Jensen}
\affiliation{Niels Bohr Institute, Danish Quantum Optics Center  QUANTOP, Copenhagen University, Blegdamsvej 17, 2100 Copenhagen, Denmark}
\author{H. Krauter}
\affiliation{Niels Bohr Institute, Danish Quantum Optics Center  QUANTOP, Copenhagen University, Blegdamsvej 17, 2100 Copenhagen, Denmark}
\author{J. J. Renema}
\affiliation{Niels Bohr Institute, Danish Quantum Optics Center  QUANTOP, Copenhagen University, Blegdamsvej 17, 2100 Copenhagen, Denmark}
\author{M. V. Balabas}
\affiliation{Niels Bohr Institute, Danish Quantum Optics Center  QUANTOP, Copenhagen University, Blegdamsvej 17, 2100 Copenhagen, Denmark}
\author{E. S. Polzik}
\affiliation{Niels Bohr Institute, Danish Quantum Optics Center  QUANTOP, Copenhagen University, Blegdamsvej 17, 2100 Copenhagen, Denmark}

\begin{abstract}
We study experimentally the fundamental limits of sensitivity of an atomic radio-frequency magnetometer.
First we apply an optimal sequence of state preparation, evolution, and the back-action evading measurement to achieve a nearly projection noise limited sensitivity. We furthermore experimentally demonstrate that Einstein-Podolsky-Rosen (EPR) entanglement of atoms generated by a measurement enhances the sensitivity to pulsed magnetic fields. We demonstrate this quantum limited sensing in a magnetometer utilizing a truly macroscopic ensemble of $1.5\cdot10^{12}$ atoms which allows us to achieve sub-femtoTesla$/\rtHz$ sensitivity.

\end{abstract}
\maketitle

Ultra-sensitive atomic magnetometry is based on the measurement of the polarization rotation of light transmitted through an ensemble of atoms placed in the magnetic field \cite{BudkerNP2007}. For $N_A$ atoms in a state with the magnetic quantum number $m_{F}=F$ along a quantization axis $x$ the collective magnetic moment (spin) of the ensemble has the length $J_{x}=FN_A$. A magnetic field along the $y$ axis
causes a rotation of $\vec{J}$ in the $x-z$ plane. Polarization of light propagating along $z$ will be rotated proportional to $J_z$ due to the Faraday effect. From a quantum mechanical point of view, this measurement is limited by quantum fluctuations (shot noise) of light, the projection noise (PN) of atoms, and the quantum backaction noise of light onto atoms. PN originates from the Heisenberg uncertainty relation $\delta J_z \cdot\delta J_y \geq J_x/2$, and corresponds to the minimal transverse spin variances $\delta J^{2}_{z,y} =J_x/2=FN_A/2$ for uncorrelated atoms in a coherent spin state \cite{WinelandPRA92}. Quantum entanglement leads to the reduction of the atomic noise below PN and hence is capable of enhancing the sensitivity of metrology and sensing as discussed theoretically in \cite{WinelandPRA92,HuelgaPRL97, AuzinshPRL04, AndrePRL04, MolmerPRA05,MabuchiPRA04,Kominis08,Lukin09}. In \cite{Leibfried2004,Roos2006} entanglement of a few ions have been used in spectroscopy. Recently proof-of-principle measurements with larger atomic ensembles, which go beyond the PN limit have been implemented in interferometry with $10^{3}$ atoms \cite{Esteve2008}, in Ramsey spectroscopy \cite{SchleierSmith09,AppelPNAS09} with up to $10^{5}$ atoms, and in Faraday spectroscopy with $10^{6}$ spin polarized cold atoms \cite{Koschorreck:}.

\begin{figure}
\centering

\includegraphics[scale=0.45]{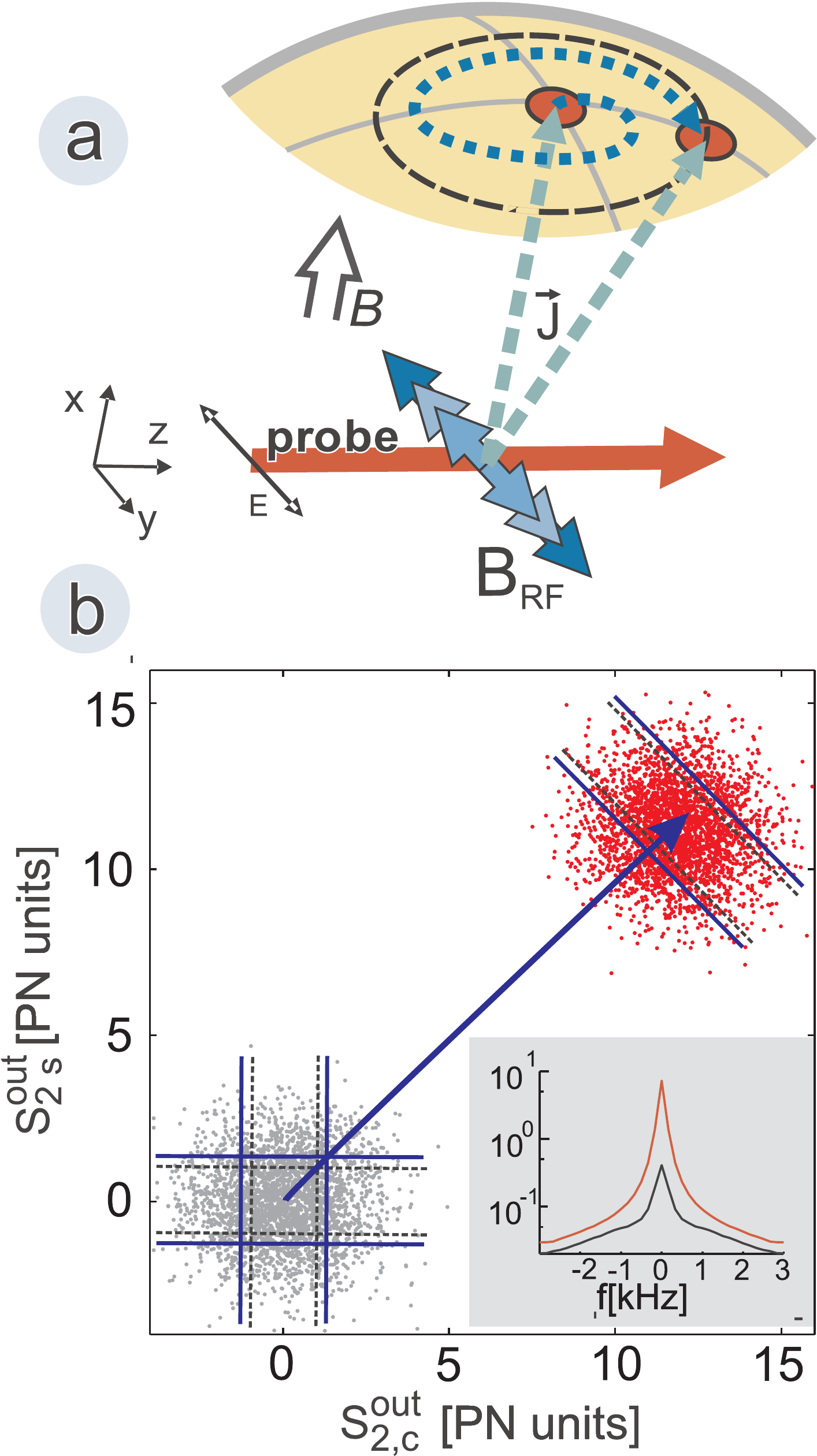} 
\caption{(a) Radio-frequency magnetometer. The atomic spin $\vec J$ precesses in crossed dc and RF magnetic fields (the blue dashed spiral).
The precessing $\vec J$ imposes an oscillating polarization rotation on the probe light.
(b) Results of a series of measurements of $\hat{S}_{2s,c}$ without $B_{RF}$ (grey points) and after a 15~msec RF pulse with $B_\text{RF}=36$~fT (red points). The solid and the dashed lines show the standard deviation of the experimental points, and of the the projection noise (PN) limited measurement, respectively. The spin displacement in the rotating frame is indicated by the dashed blue arrow. (Inset) Power spectrum of the photocurrent (arb. units, centered at $\Omega$)
 The narrow peak of the red/black spectrum is the atomic spin signal/projection noise.
 The broad part of the spectrum is the shot noise of light with the width set by the detection bandwidth.}
\label{fig:cloudplot1}
\end{figure}

\begin{figure}
\centering
\includegraphics[width=0.4\textwidth]{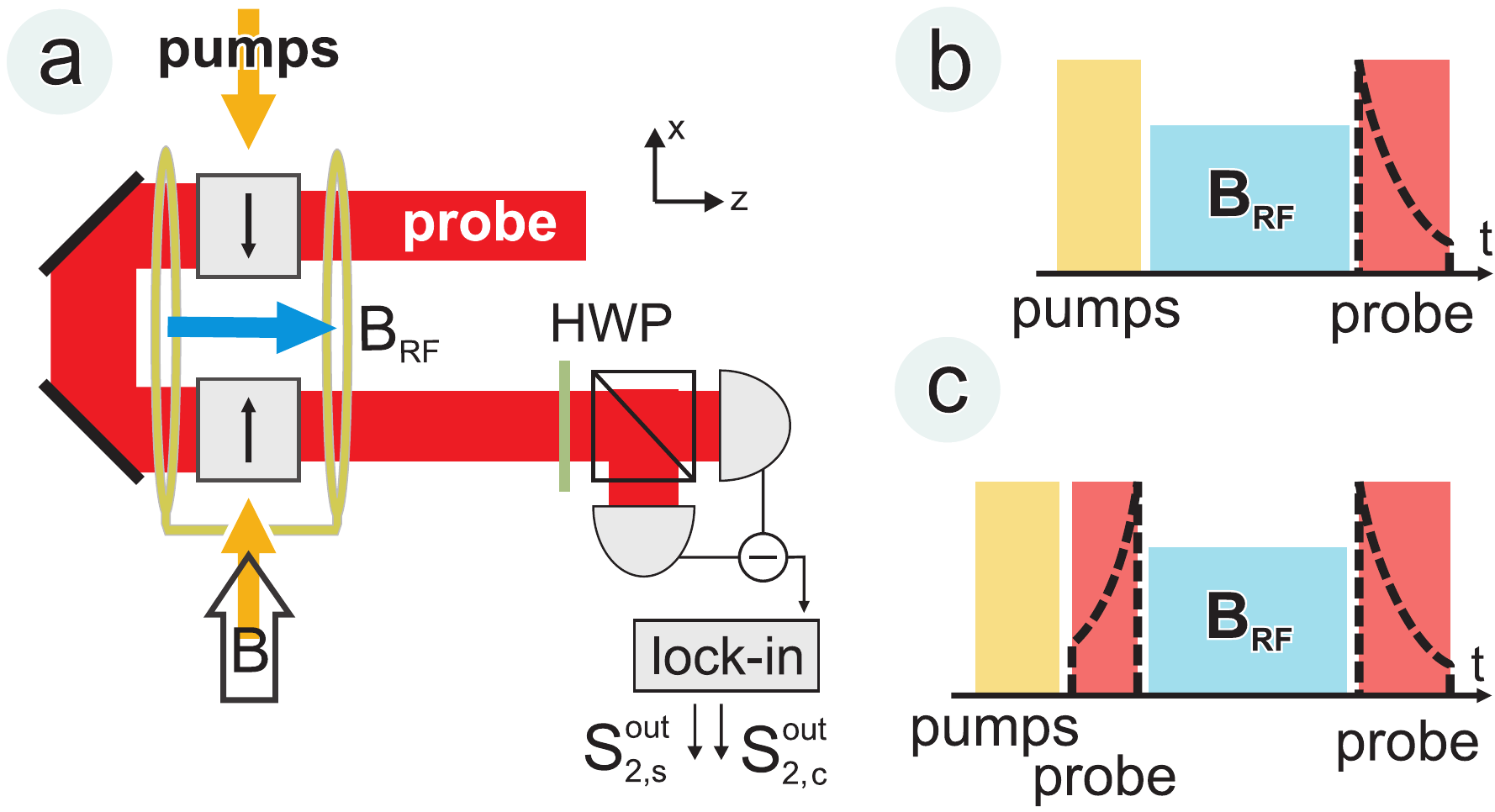}	 
\caption{(a) Schematic of the experimental setup. Cubic $22$mm paraffin coated cells filled with caesium are placed in the dc field $B$ in two separate magnetic shields. The atoms are optically pumped so that the directions of the collective spins in the two cells (black arrows) are opposite. A pulse of $B_{RF}$ at the frequency $\Omega$ is applied orthogonally to the $B$ field. The polarization rotation of the top-hat shaped probe beam pulse (diameter $21$mm)
is detected by  two detectors (HWP - half wave plate).
The lock-in amplifier measures the $\cos\Omega t$ and $\sin\Omega t$ components of the photocurrent.
(b) Pulse sequence for projection noise limited magnetometry. The temporal mode function for the probe is shown with a dashed black curve. (c) Pulse sequence and temporal modes for entanglement-assisted magnetometry.}
\label{fig:setup}
\end{figure}

 In this Letter we demonstrate PN limited and entanglement-assisted measurement of a radio-frequency (RF) magnetic field by an atomic caesium vapour magnetometer.
In the magnetometer $\vec{J}$ precesses at the Larmor frequency $\Omega/2\pi=322$kHz around a dc field $B=0.92$G applied along the $x$ axis and an RF field with the frequency $\Omega$ is applied in the $y-z$ plane (Fig.~\ref{fig:cloudplot1}a). The magnetometer (Fig.~\ref{fig:setup}a) detects an RF pulse with a constant amplitude $B_\text{RF}$ and duration $\tau$ (Fourier limited full width half maximum bandwidth $\delta=0.88 \tau^{-1} \approx \tau^{-1}$). The mean value of the projection of the atomic spin on the $y-z$ plane in the rotating frame after the RF pulse is
$ \Gamma B_\text{RF}J_{x} T_{2}[1-\exp(- \tau /T_{2})] /2$. Here $T_2$ is the spin decoherence time during the RF pulse and $\Gamma = \Omega/B= 2.2\cdot 10^{10}$rad/(sec$\cdot$Tesla) for caesium. Equating the mean value to the PN uncertainty we get for the minimal detectable field under the PN limited measurement
\begin{equation}
B_\text{min}=\left[\Gamma \sqrt{F N_A/2}\cdot T_2\left( 1-e^{-\tau/T_2} \right) \right]^{-1}\label{eq:Bmin}.
 \end{equation}

 The PN-limited sensitivity to the magnetic field is then $B_\text{min}\sqrt{\tau}$, and equals the standard deviation of the measured magnetic field which can be achieved by using repeated measurements with a total duration of 1 second.
 The best sensitivity to the $B_{RF}$ with a given $\delta$ is achieved with the narrow atomic bandwidth: $\delta = \tau^{-1} \geq {T_2}^{-1}$.
 A long $T_2$ also helps to take advantage of the entanglement of atoms. Entangled states are fragile and have a shorter lifetime $T_{2E}<T_2$. As we demonstrate here, under the condition $\delta^{-1} <T_{2E}<T_2$, that is for broadband RF fields, entanglement improves the sensitivity. Similar conclusions have been reached theoretically for atomic clocks in \cite{HuelgaPRL97, AndrePRL04} and for dc magnetometry in \cite{AuzinshPRL04}.

In our experiment a long magnetometer coherence time $T_2>30$msec is achieved by using paraffin coated caesium cells at around room temperature \cite{Sherson2006}, and by the time resolved quantum spin state preparation (optical pumping), evolution, and measurement.
In this way $T_2$ is not reduced by the optical pumping and/or measurement-induced decoherence during the time when magnetic field is applied.

PN limited sensitivity requires, besides elimination of the technical noise, the reduction of the back action noise of the measurement and of the shot noise of the probe light.

\begin{figure}
\centering
\includegraphics[width=0.45\textwidth]{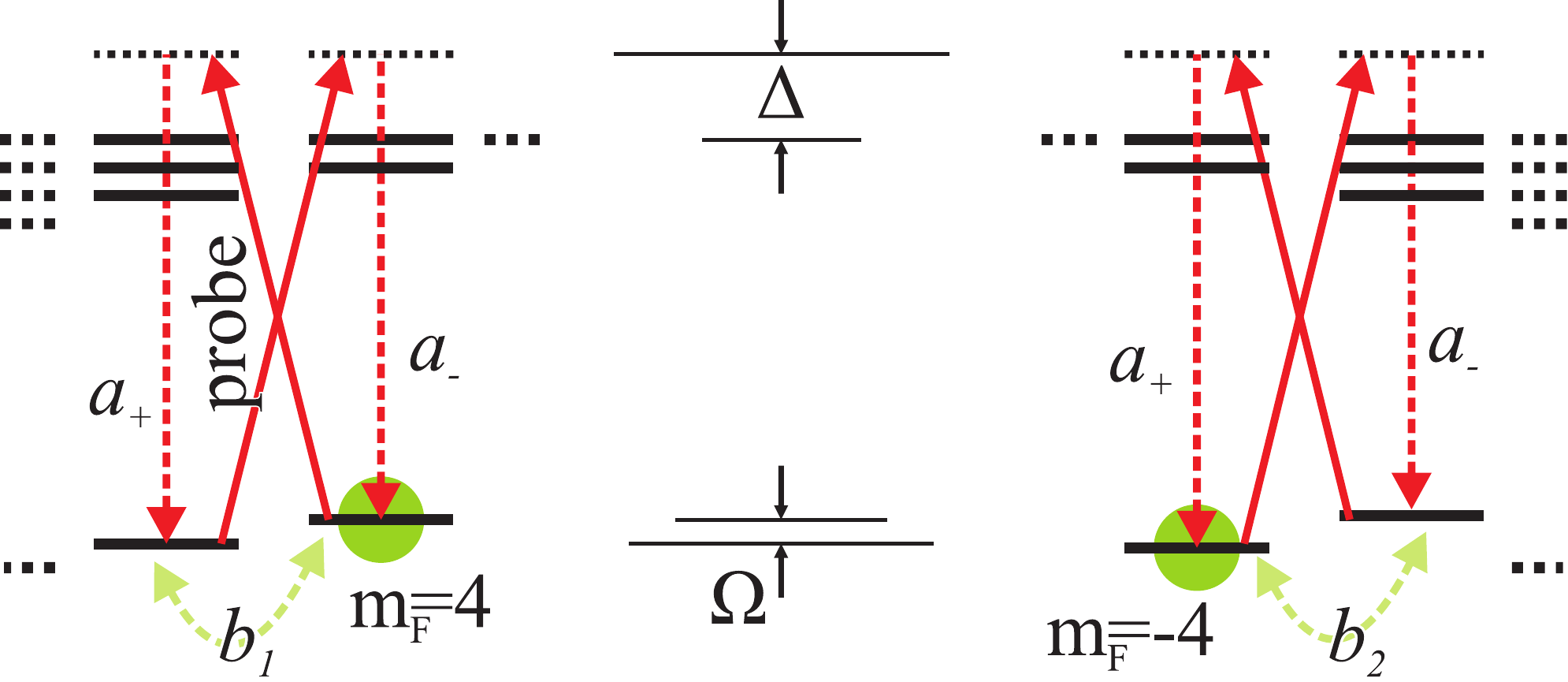} 
\caption{Level scheme and optical fields used in the magnetometer. The input probe light polarized orthogonally to the magnetic field (solid arrows) is blue detuned by $\Delta=850$MHz from the $F=4\rightarrow F=5$ hyperfine component of the $D2$ line at $852$nm. Forward scattering of photons into modes $a_{\pm}$ leads to polarization rotation and swapping interaction described in the text. The two diagrams correspond to the two oppositely polarized cells using which leads to cancelation of the backaction of light onto atoms, as explained in the text.}
\label{fig:levels}
\end{figure}

The back action noise comes from the Stark shift imposed by quantum fluctuations of light polarization on $J^{lab}_y$ in the laboratory frame when $J^{lab}_z$ is measured. When atoms are exposed to an oscillating magnetic field $J^{lab}_z$ and $J^{lab}_y$ experience Larmor precession, and hence both of them accumulate the back action noise \cite{Julsgaard2001,SavukovPRL05}. As proposed in \cite{Julsgaard2001} the effect of this noise can be canceled if two atomic ensembles with orientations $J_x$ and $-J_x$ are used. Fig.~\ref{fig:setup}a presents the sketch of the magnetometer layout
which is used for the back action evading measurement of the $B_{RF}$, and Fig.~\ref{fig:levels} shows the level schemes for the two atomic ensembles.

To read out the atomic spin, which carries information about the RF magnetic field, we utilize polarized light. The  quantum state of light is described by the Stokes operators $\hat S_1$, $\hat S_2$ and $\hat S_3$. The relevant observables are the cos/sin $\Omega$ Fourier components of $\hat S_2$ integrated over the pulse duration $T$ with a suitable exponential modefunction $\hat{S}_{2c}\propto \int_0^T \hat{S}_{2}(t) \cos (\Omega t) e^{\pm \gamma_{\text{tot}}t}$.
$\hat{S}_{2s}$ is similarly defined by replacing cosine with sine.
The operators $\hat{S}_{2c}$ and $\hat{S}_{2s}$ are experimentally measured by homodyning and lock-in detection (Fig.~\ref{fig:setup}a).

The theory of interaction between the probe light and a vapor of spin polarized alkali atoms has been developed in \cite{KupriyanovPRA05,Wasilewski09}.
After the probe has interacted with the two atomic ensembles, the output Stokes operator (normalized for a coherent state as $\var{S}=\Phi /2$, with $\Phi$ being the photon number per pulse) is given by \cite{Wasilewski09}
\begin{equation}
\hat{S}^{out}_{2c}=\hat{S}^{in}_{2c}\sqrt{1-\xi^{2}\kappa^{2}}+\kappa \sqrt{\frac{\Phi}{2FN_A}}(\hat{J}^{in}_{z1} +\hat{J}^{in}_{z2}) \label{eq:inout},
\end{equation}
with the equation for $\hat{S}^{out}_{2s}$ obtained by the replacement $\hat{J}^{in}_{z}\rightarrow \hat{J}^{in}_{y}$.
The output Stokes operator is defined with an exponentially falling mode function $e^{-\gamma_{\text{tot}}t}$, and the input Stokes operator is defined with a rising mode $e^{+\gamma_{\text{tot}}t}$ \cite{Wasilewski09}, where $\gamma_{\text{tot}}$ is the decay rate of the transverse atomic spin components.
$\kappa=\xi^{-1} \sqrt{1-\exp(-2\gamma_\text{swap} T)}$ is the light-atom coupling constant, $\gamma_\text{swap}$ is proportional to intensity of light and density of atoms, and
$\xi=\sqrt{{14a_2}/{a_1}}$, where $a_2$ and $a_1$ are the tensor and vector polarizabilities. For our probe detuning $\Delta=850$MHz, $\xi^2=1/6.3$. In the two-cell setup $\hat{S}_{2c}$ and $\hat{S}_{2s}$ contain information about the commuting rotating frame operators $\hat J^{in}_{y+}=\hat{J}^{in}_{y1} +\hat{J}^{in}_{y
 2}$ and $\hat J^{in}_{z+}=\hat{J}^{in}_{z1} +\hat{J}^{in}_{z2}$ simultaneously, which are displaced by $B_{RF}$.

 The atomic spin operator $J_{z+}$ after the interaction is given by
\begin{equation}\label{Eq:Joutz+}
\hat J^{out}_{z+} =(\hat J^{in}_{z1} +\hat J^{in}_{z2}) \sqrt{1-\xi^{2}\kappa^{2}}
-\xi^2 \kappa  \sqrt{\frac{2FN_A}{\Phi}}\hat{S}^{in}_{2c},
\end{equation}
In case of $\gamma_\text{swap} T\ll 1$, corresponding to either rather large probe detuning $\Delta$, or to a relatively small photon number $\Phi$, as in \cite{Julsgaard2001,Julsgaard2004}, Eq.~\eqref{eq:inout} and \eqref{Eq:Joutz+} reduce to the quantum nondemolition (QND) measurement, where $J_{y+}$ and $J_{z+}$ are conserved. Here we implement a strong measurement limit, $\gamma_\text{swap} T\approx 1$ in which the light and atoms nearly swap their quantum states. Indeed under this condition first terms in Eq.~\eqref{eq:inout}) and \eqref{Eq:Joutz+} are strongly suppressed. The exponential suppression of the probe shot noise contribution to the signal with time and with the optical depth $d$ in Eq.~\eqref{eq:inout}) allows to approach the PN limited sensitivity faster than in the QND measurement case where this contribution would stay constant.

The PN limit can be approached if the total rate of the decay of the atomic coherence  $\gamma_\text{tot}$ is dominated by the coherent swap rate $\gamma_\text{swap}$. In the experiment we obtained $\gamma_\text{swap}=0.43$ms$^{-1}$ and $\gamma_\text{tot}=0.50$ms$^{-1}$ with a $15$mW probe detuned by $850$MHz and $N_A=2\cdot7.2(7) \cdot 10^{11}$ corresponding to the effective resonant optical depth $d=75$. The residual decoherence rate due to  the spontaneous emission rate $\eta_{1}$,
collisions, and magnetic dephasing is therefore $0.07$ms$^{-1}$. Note that $\gamma_\text{swap}\sim\xi^{2}d\cdot \eta_{1}\gg \eta_{1}$.  The probe losses are dominated by reflection on cell walls and on detection optics, since its absorption in the gas is negligible.

Raw results for a series of measurements of $\hat{S}^{out}_{2c,s}$ obtained at $32^\circ$C corresponding to $N_A=2\cdot7.2(7) \cdot 10^{11}$ are presented in Fig.~\ref{fig:cloudplot1}b. $N_A$ is monitored via $J_{x}=4N_{A}$ measured by the Faraday rotation angle $19^\circ$ of an auxiliary probe beam sent along the $x$ axis, and from the degree of spin polarization $>98\%$, as determined from the magneto-optical resonance \cite{Julsgaard:2004a}. The experimental sensitivity is $B_{RF}\cdot \sqrt{\tau}/SNR=3.6(4) \cdot 10^{-16}$Tesla$/\rtHz$, where the signal to noise ratio SNR$=12.3$ is found from the data in Fig.~\ref{fig:cloudplot1}b as the ratio of the mean to the standard deviation of the data (red points) obtained for $B_{RF}=36(3)$fT applied during $\tau=15$msec via a calibrated RF coil. The optimal temporal mode function for this measurement has $\gamma_{opt}=2\gamma_{tot}=1.0$ms$^{-1}$. This sensitivity is $\sim30\%$ above the theoretical PN limited sensitivity $2.7(5)\cdot 10^{-16}$Tesla$/\rtHz$ found from (Eq.~\eqref{eq:Bmin}) using the measured value of $T_{2}=32$ms. Using this sensitivity we can calibrate the values in Fig. 1b in PN units. The difference between the two sensitivities is due to the residual contribution of the shot noise of the probe (first term in Eq.~\eqref{eq:inout}, the decay of the atomic state during the RF pulse and the classical fluctuations of the atomic spins.

The experimental sensitivity $B_{RF}\cdot \sqrt{T}/SNR=4.2(8) \cdot 10^{-16}$Tesla$/\rtHz$ calculated using the full measurement cycle time $T$ including the duration of optical pumping ($6$ms), probing ($1.5$ms) and $\tau=22$ms (Fourier limited bandwidth $\sim 40$Hz) approaches the best to-date atomic RF magnetometry sensitivity \cite{LeeAPL06} obtained with $10^{4}$ times more atoms. This is to be expected since PN limited magnetometry yields the best possible sensitivity per atom achievable without entanglement.

\begin{figure}
\centering
\includegraphics[width=0.45\textwidth]{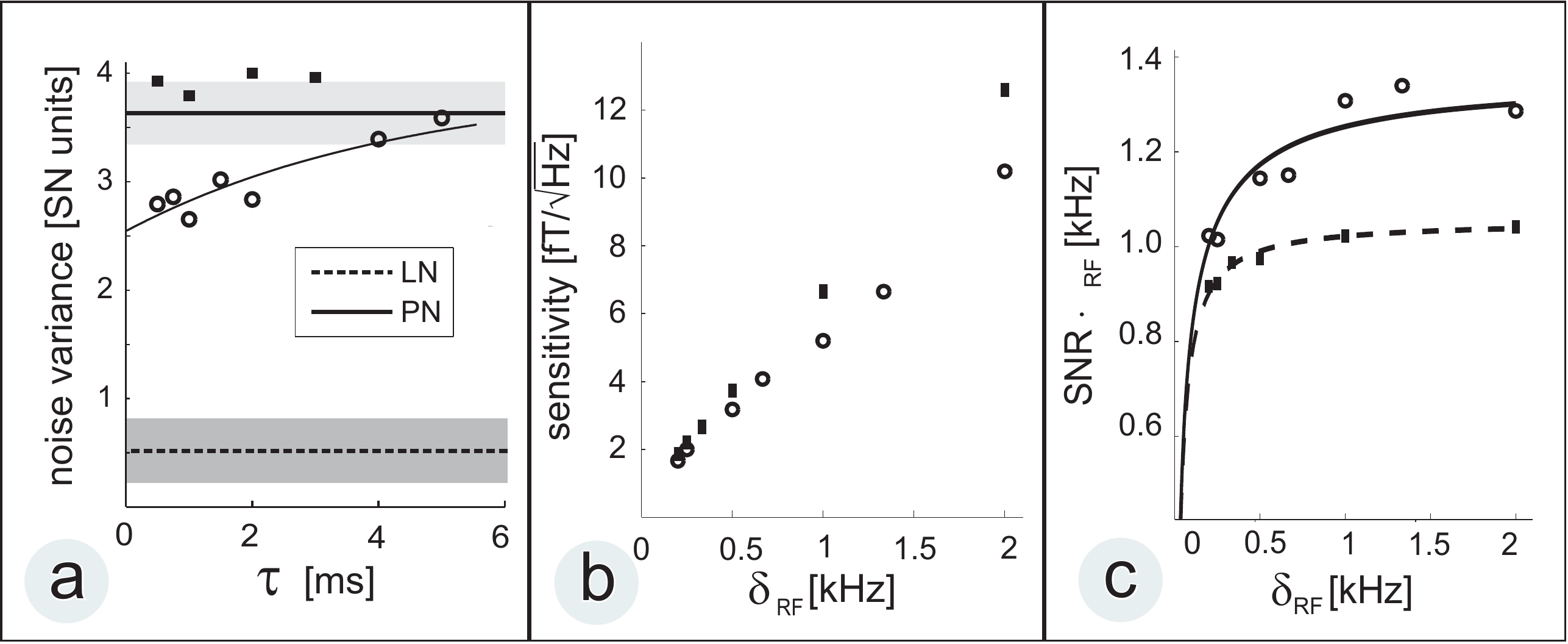} 
\caption{Entanglement assisted magnetometry results. (a) Magnetometer read out noise in units of shot noise of light. The dashed line at $0.5$ is the contribution of the probe light noise(LN).  The average noise level of the initial spin state (squares) is $1.10(8)$ in projection noise (PN) units. The read out noise corresponding to the PN level is shown as the solid horizontal line. Grey bands have the width equal to twice the standard deviation. The magnetometer noise for the entangled state vs the delay between the two probe pulses is shown as circles. The fit yields the entanglement lifetime $T_{2E}=4$msec. (b) Sensitivity of the entanglement assisted magnetometer (circles) and the sensitivity in the absence of entanglement (squares) vs the RF pulse bandwidth (inverse pulse duration). The $B_\text{RF}=36$~fT pulse is applied to $N_{A}=3.6(3)\cdot10^{11}$ atoms. The duration of the first/second probe pulse is $2/3$msec, the duration of the pumping cycle $6$msec. (c) Signal/noise ratio times the RF bandwidth for the magnetometer in the presence (circles) and in the absence (squares) of the entanglement.}
\label{fig:ent1}
\end{figure}

We now turn to the entanglement-assisted magnetometry. As first demonstrated in \cite{Julsgaard2001} the measurement of the Stokes operators $\hat{S}^{out}_{2c,s}$  can generate the state of atoms in the two cells which fulfils the Einstein---Podolsky---Rosen (EPR) entanglement condition for two atomic ensembles with macroscopic spins $J_x$ and $-J_x$ : $\Sigma_\text{EPR}=\left[\var{\hat{J}^{in}_{y1} +\hat{J}^{in}_{y2}} + \var{\hat{J}^{in}_{z1} +\hat{J}^{in}_{z2}}\right]/{2J_x}<1$.  This inequality means that the atomic spin noise which enters in Eq.~\eqref{eq:inout} is suppressed below the PN level corresponding to the coherent spin state. Entanglement can be visualized as correlations between the grey and red points shown in Fig.~\ref{fig:cloudplot1}b. The degree of entanglement and the sensitivity are optimized by choosing suitable rising/falling probe modes $e^{\pm\gamma t}$ for the first/second probe modes shown in (Fig. \ref{fig:setup}c). In order to reduce the contribution of the technical (classical) fluctuations of the spins we perform the entanglement-assisted measurements at room temperature with $N_A=3.6\cdot 10^{11}$ per cell.

In order to demonstrate entanglement of atoms, we need to calibrate the PN level. Above we have calculated the atomic noise in PN units with $20\%$ uncertainty using (Eq.~\eqref{eq:Bmin}). However this uncertainty coming from the uncertainties of $N_{A}$ and $B_{RF}$ is too high and we therefore apply the following calibration method.
The collective spin operator $\hat J_{y-} =\hat J_{y1} -\hat J_{y2}$ after the interaction with a probe pulse is given by \cite{Wasilewski09}: $\hat J^{out}_{y-} =
(\hat J^{in}_{y1} -\hat J^{in}_{y2}) \sqrt{1-\xi^{2}\kappa^{2}}
+\kappa \sqrt{\frac{2FN_A}{\Phi}}\hat{S}^{in}_{3c}$.
Using an electro-optical modulator we create a certain average value of $\hat S^{in}_{3c}$ and then calibrate it in units of shot noise $\sqrt{\Phi/2}$ by homodyne detection. We determine $\kappa^2$ for a certain light power $P\approx8$mW by first sending a 1msec pulse with $\avg{\hat S^{in}_{3c}}=\sqrt{\Phi/2}$, which is mapped on the atoms. We then flip the sign of $\hat J_{y2}$ and $\hat J_{z2}$ by momentarily increasing the dc magnetic field and send a second 1msec light pulse.  Now $\avg{\hat J_{y-}}$ is mapped on $\avg{\hat S^{out}_{2s}}$, following Eq.~\eqref{eq:inout}, and the measurement on the second pulse yields $\avg{\hat S^{out}_{2s}}=\kappa^2$. Using $\kappa^{2}$ and the detection efficiency $\eta=0.8$ we can calculate the atomic noise in projection noise units from the measured noise $\var{\hat{S}_{2c,s}^{out}}$ using  Eq.~\eqref{eq:inout}. With a known value of $\kappa$, we can now calibrate the atomic displacement $(\avg{\hat J_y},\avg{\hat J_z})$ caused by a particular $B_{RF}$ in units of PN as follows. The mean value of the homodyne signal is $\avg{\hat {\tilde S}^{out}_2}=\kappa\sqrt\eta \sqrt{\Phi/2FN_A} (\avg{\hat J_{y1}+\hat J_{y2}}+i\avg{\hat J_{z1}-\hat J_{z2}})$. From this expression the atomic displacements in units of PN $\sqrt{F N_A}$ can be found using the values of $\kappa, \eta$ and the shot noise $\Phi/2$.
Once we know the atomic displacement in units of PN for the RF pulse with a certain duration and amplitude we can utilize this to find $\kappa$ and hence the PN level for probe pulses with any light power (most of the measurements were done with a probe power $P\approx14$mW) from the mean value of the homodyne signal corresponding to the RF pulse with the same duration and amplitude.

Entanglement is generated in our magnetometer by the entangling light pulse  applied before the RF pulse in the two-pulse time sequence (Fig.~\ref{fig:setup}c). Fig.~\ref{fig:ent1}a shows the magnetometer noise $\var{\hat{S}^{out}_{2}}$ including the light noise and the atomic noise contributions in units of shot noise of the $15$mW probe.
as a function of the RF pulse duration for entangled atoms and for atoms in the initial state. For these measurements $\kappa^{2}=3.1$ which is consistent with Fig.~\ref{fig:ent1}a since, according to Eq.~\eqref{eq:inout}, for PN limited measurement $\kappa^{2}$ is equal to the total noise $3.6$ less the light noise $0.5$.
Knowing $\kappa^{2}$ we find the initial noise level in PN units to be $1.10(8)$.
The best noise reduction below PN of about $-1.5$dB (-30\%) is obtained for short RF pulses corresponding to the RF bandwidth $\delta_{RF}\geq 1$ kHz. Fig.~\ref{fig:ent1}b illustrates the improvement in the sensitivity $B_{RF}\cdot \sqrt{T}/SNR$ with entangled atoms compared to the sensitivity obtained with atoms in the initial state.
Fig.~\ref{fig:ent1}c shows that entanglement improves the product of the S/N ratio times the bandwidth $\delta_{RF}$ by a constant factor for RF pulses with $\delta_{RF}$ much greater than the inverse entanglement lifetime $(\pi T_{2E})^{-1}=70$Hz.

To the best of our knowledge, our results present entanglement assisted metrology with the highest to-date number of atoms. This results in the absolute sensitivity at the sub-femtoTesla$/\rtHz$ level which is comparable to the sensitivity of the state-of-the-art atomic magnetometer \cite{LeeAPL06} operating with orders of magnitude more atoms.
The two-cell setup can also serve as the magnetic gradient-meter if the direction of the probe in the second cell is flipped or if the RF pulse is applied to one cell only. Increasing the size of the cells to a $5$~cm cube should yield the sensitivity of the order of $5\cdot 10^{-17}$T$/\rtHz$ which approaches the sensitivity of the best superconducting SQUID magnetometer \cite{Seton05}. The degree of entanglement $\Sigma_\text{EPR}$ can, in principle, reach the ratio of the tensor to vector polarizabilities $\xi^{2}$ which is $0.16$ or $-8$dB for our experiment, and can be even higher for a farther detuned probe. This limit has not been achieved in the present experiment due to various decoherence effects, including spontaneous emission and collisions. Increasing the size of the cells may help to reduce some of those effects since a larger optical depth will then be achieved for a given density of atoms.

This research was supported by EU grants QAP, COMPAS and Q-ESSENCE.


\end{document}